\documentclass[preprint2]{aastex}
\usepackage{amsmath}
\allowdisplaybreaks
\shorttitle{Axisymmetric Limit of PN Dedekind Ellipsoids}
\shortauthors{G\"urlebeck and Petroff}
\renewcommand{\d}{\text{d}}

\begin{document}

\title{The Axisymmetric Case for the Post-Newtonian Dedekind Ellipsoids}

\author{Norman G\"urlebeck}
\affil{Institute of Theoretical Physics, Charles University, Prague, Czech Republic}
\email{norman.guerlebeck@gmail.com}

\author{David Petroff}
\affil{Institute of Theoretical Physics, Friedrich-Schiller University, Jena, Germany}
\email{D.Petroff@tpi.uni-jena.de}

\begin{abstract}
We consider the post-Newtonian approximation for the Dedekind
ellipsoids in the case of axisymmetry. The approach taken by \citet{CE74, CE78_CE78}
excludes the possibility of finding a uniformly rotating
(deformed) spheroid in the axially symmetric limit, though the
solution exists at the point of axisymmetry. We consider an extension to their
work that permits the possibility of such a limit.
\end{abstract}


\section{Introduction}

The Dedekind tri-axial ellipsoids are an example of non-axisymmetric, but stationary
solutions within Newtonian gravity. Due to internal motions, they are, in fact, stationary in
an inertial frame. When addressing the question of whether or not stationary, but non-axisymmetric
solutions are possible within General Relativity, this property makes the Dedekind ellipsoids a
natural choice upon which to base one's considerations. It was, in part, with this question in
mind that \citet{CE74, CE78_CE78} turned their attentions to the post-Newtonian (PN)
approximation of the Dedekind ellipsoids. In a paper from the same series,
\citet{Chandrasekhar67c_C67} had already
considered the axisymmetric limit of the PN Jacobi ellipsoids at length and was able to show
that it coincides with a certain PN Maclaurin spheroid (just as their
Newtonian counterparts coincide at the point of bifurcation). This is related to the fact that
the PN figures were chosen to rotate uniformly. On the other hand, the PN velocity field chosen
in \cite{CE78_CE78} excludes the possibility of uniform rotation in the axisymmetric
{\sl limit} although it is possible in the axisymmetric {\sl case}.
This restriction seems neither natural nor advisable in the context of trying to settle the
question as to the existence of relativistic, non-axisymmetric, stationary solutions. The
na{\"{\i}}ve expectation is that the axisymmetric PN Dedekind ellipsoids contain the PN
Maclaurin spheroids in the axisymmetric limit (up to arbitrary order).
\par

In this article, we begin in \S~\ref{axisymm_point} by examining the axisymmetric case of 
a generalization to the solution presented in \citet{CE78_CE78}. We proceed in
\S~\ref{axisymm_limit} to consider a (continuous) limit to axisymmetry. 
In \S~\ref{discussion} the connection to the post-Newtonian Maclaurin spheroids is examined.

\section{The Axisymmetric Solution of a Generalization to Chandrasekhar and Elbert's Paper}\label{axisymm_point}

We consider a generalization of the PN Dedekind ellipsoids presented in \citet{CE78_CE78} 
(referred to from here on in as CE78) in which we add post-Newtonian terms to the velocity. We
comply with the notation used in CE78 and refer the reader to the definitions there for the
various quantities. The post-Newtonian contributions to the velocity, which we introduce here are
\begin{align}\label{velocity}
 \begin{split}
  \delta v_1 &= a_1^2 w_1 x_2 + (q_1+q)x_1^2 x_2 + r_1 x_2^3 + t_1 x_2 x_3^2\\
  \delta v_2 &= a_2^2 w_2 x_1 + (q_2-q)x_1  x_2^2 + r_2 x_1^3 + t_2 x_1 x_3^2\\
  \delta v_3 &= q_3 x_1 x_2 x_3,
 \end{split}
\end{align}
where the terms with $w_1$ and $w_2$ have been added for reasons that will be made
clear when we discuss the solution. Note that we could eliminate one constant by
introducing variables to denote $q_1+q$ and $q_2-q$, but choose instead to retain
the notation in CE78.%
\footnote{The three-velocity $v^i$ in CE78 does not refer to the spatial components of the
four-velocity $u^\alpha=\d x^\alpha/\d\tau$, but is instead defined as $v^i=
\d x^i/\d t= u^i/u^0$.}
\par

The Newtonian ellipsoid is characterized by the semi-major axes $a_1\ge a_2\ge a_3$.
Let us assume for the moment that, as in the Newtonian setting, the axisymmetric
case is obtained by considering $a_2=a_1$, an assumption that will be verified
shortly. In this case, the index `2' in the index symbols $A_{ijk\ldots}$ and $B_{ijk\ldots}$ used in CE78 and
discussed at length in \S~21 of \citet{Chandrasekhar87} can be replaced by `1' as is evident
from the definitions. Using the relations given in that book, it is possible to reduce all the index symbols to $A_1$ and $A_2$. At the point $a_2=a_1$, the value for $A_1$ (and thus $A_2$)
is given by (36) in \S~17 of \citet{Chandrasekhar87}. Furthermore, (2) from \citet{CE74} shows us that
\begin{align}\label{Q}
 Q_2\stackrel{a}{=} -Q_1,
\end{align}
where we define the symbol $\stackrel{a}{=}$ to mean that the
expression is evaluated at the point $a_2=a_1$, i.e.
\begin{align}
 \left.C\right|_{a_2=a_1} = \left.D\right|_{a_2=a_1}\ \text{is denoted by}\ C\stackrel{a}{=} D.
\end{align}
The value for $a_3$ can be found from the equation
\begin{align}\label{a3_equation}
 a_1^2 a_2^2 A_{12}=a_3^2A_3,
\end{align}
which holds for the Dedekind (and Jacobi) ellipsoids, and gives the value
\begin{align}
 \frac{a_3}{a_1}\stackrel{a}{=} 0.5827241661\ldots.
\end{align}
Throughout this paper, $a_3$ is to be understood as a function of $a_1$ and $a_2$,
given by \eqref{a3_equation}.
\par

We can now consider the integrability conditions for the pressure
and the continuity equation. We again follow CE78 and shall
refer to the equation numbers there by adding a prime. It turns
out that (38') (of CE78) remains unchanged despite the
modification to the velocity, so that we find 
\begin{align}
 q_3\stackrel{a}{=} 0
\end{align}
and then from (24') that
\begin{align}\label{q2}
 q_2\stackrel{a}{=} -q_1. 
\end{align}
Equation~(28') is identically fulfilled for $a_2=a_1$, meaning that $q_1$ is left
undetermined, in contrast to the general case.
\par

With the changes to the velocity, equations (30') and (31')
gain the additional terms $(a_1^2Q_2w_1+a_2^2Q_1w_2)x_1$ and $(a_1^2Q_2w_1+a_2^2Q_1w_2)x_2$
respectively. Equations (32')--(38') remain unchanged. Equation (32') yields 
\begin{align}\label{r2}
 r_2\stackrel{a}{=} -r_1
\end{align}
and (37') gives 
\begin{align} \label{t2}
 t_2\stackrel{a}{=} t_1
\end{align}
(we shall see shortly that each $t_i$ becomes zero). There are additional
terms in (39') corresponding to adding $-(a_1^2Q_2w_1+a_2^2Q_1w_2)/2=a_2^2Q_1(w_1-w_2)/2$ both to%
\footnote{We use the superscripts `67' and `78' to distinguish the quantities defined in
\citet{Chandrasekhar67c_C67} from those in \citet{CE78_CE78}.}
$\alpha_1^{78}$ and $\alpha_2^{78}$.
\par

Requiring for the new velocity that its normal component vanish on the surface leads
to a change in (50') and thus the resulting equations (52')--(56') by which the terms with $S_1-S_2$ are modified. They now become
\begin{align}
 S_1- S_2-\frac{w_1+w_2}{2Q_1}.
\end{align}
Using equations \eqref{Q}, \eqref{q2} and \eqref{r2}, we can subtract
equation (54') from (55') in CE78 to arrive at
\begin{align}\label{q_minus_r1}
 q+q_1-r_1\stackrel{a}{=} \tfrac43 Q_1(4S_3+S_4). 
\end{align}
\par

Next we turn our attention to the system of equations (58') from CE78%
\footnote{Please note that we have been unable to reproduce the values from Table~1' in CE78
that result from solving (58'). A detailed discussion can be found in Appendix~\ref{details}.\label{different_solution}}.
In the case being considered here, the first of these equations becomes
\begin{align}
 0 &= Q_{11}^{78} + \frac{a_2^4}{a_1^4}Q_{22}^{78} - \frac{a_2^2}{a_1^2}Q_{12}^{78} \nonumber\\
   &\stackrel{a}{=} P_{11}^{78} + P_{22}^{78} - P_{12}^{78}
            - \frac{2a_3^2}{3a_1^2}A_3(4S_3+S_4)\nonumber\\
   &= \alpha_{11}^{78} + \alpha_{22}^{78} - \alpha_{12}^{78} + Q_1(q-r_1)\nonumber\\
  &\quad - \frac{2a_3^2}{3a_1^2}A_3(4S_3+S_4)\nonumber\\
  &\quad + \sum_{i=3}^5 S_i\left(u_{11}^{(i)}+u_{22}^{(i)}-u_{12}^{(i)}\right)\nonumber\\
  \begin{split}
  &=\tfrac23 Q_1^2(4S_3+S_4) - \frac{2a_3^2}{3a_1^2}A_3(4S_3+S_4)\\
  &\quad+ \tfrac23a_1^6A_{1111}(4S_3+S_4),
  \end{split}
\end{align}
where the values for the $\alpha$'s and their axisymmetric limits can be found in
Appendix~\ref{details}, the $u$'s are given in \citet{Chandrasekhar67c_C67} (C67b)%
\footnote{As mentioned in C67b, the $u$'s belonging to the displacements
$\boldsymbol{\xi}^{(4)}$ and $\boldsymbol{\xi}^{(5)}$ are generated by cyclically
permuting the indices. The precise meaning is best understood via the example
that $u_1^{(4)}$ can be generated from $u_3^{(3)}=-\frac12 a_1^2(a_2^2B_{123}-a_1^2B_{113})$
and becomes $u_1^{(4)}=-\frac12 a_2^2(a_3^2B_{231}-a_2^2B_{221})$.}
equations (72) and (73) and where we made use of \eqref{q_minus_r1} from the current paper. The unique solution to this equation is
\begin{align}\label{S4}
 S_4\stackrel{a}{=} -4S_3
\end{align}
as it is for the analogue equation (100) of C67b despite the
fact that the term with $q-r_1$ is absent there. With the result \eqref{S4},
\eqref{q_minus_r1} becomes
\begin{align}\label{q_plus_q1}
 q+q_1-r_1\stackrel{a}{=} 0
\end{align}
and for equation (53') from CE78%
\footnote{\label{footnote_typo} 
In (53') of CE78, the factor $Q_1$ is missing from the term with $(S_1-S_2)$.},
or equivalently the sum of (54') and (55'), we find
\begin{align}\label{S1_minus_S2}
 S_1 - S_2\stackrel{a}{=} -\frac53 S_3.
\end{align}
\par

The third minus the second of equations (58') is the analogue of equation (101)
in C67b and is in fact precisely the same equation despite the different
definitions for $P_{ij}$:
\begin{align}
 0 &= Q_{11}^{78} - \frac{a_2^4}{a_1^4}Q_{22}^{78} 
      - \frac{a_3^2}{a_1^2}Q_{31}^{78}-\frac{a_2^2a_3^2}{a_1^4}Q_{23}^{78} \nonumber\\
   &\stackrel{a}{=} P_{11}^{78} - P_{22}^{78} - \frac{a_3^2}{a_1^2}(P_{31}^{78}
          -P_{23}^{78}) \nonumber\\
   &\quad - 2\frac{a_3^2}{a_1^4} A_3\left(\tfrac{17}3a_1^2 S_3 +a_3^2 S_5\right) \nonumber\\
   &= \alpha_{11}^{78} - \alpha_{22}^{78} - \frac{a_3^2}{a_1^2}(\alpha_{31}^{78}
        -\alpha_{23}^{78}) \nonumber\\
   &\quad - 2\frac{a_3^2}{a_1^4} A_3\left(\tfrac{17}3a_1^2 S_3 +a_3^2 S_5\right) \nonumber\\
   &\quad + \sum_{i=3}^5 S_i\left[\left(u_{11}^{(i)}-u_{22}^{(i)}\right)
          -\frac{a_3^2}{a_1^2}\left(u_{31}^{(i)}-u_{23}^{(i)}\right)\right]\nonumber\\
   \begin{split}
    &= - 2\frac{a_3^2}{a_1^4} A_3\left(\tfrac{17}3a_1^2 S_3 +a_3^2 S_5\right) \\
    &\quad + \left(a_1^2A_{111}+6a_3^2A_{113}-7a_3^2A_{1113}\right)\\
    &\quad \times\left(\tfrac{17}3a_1^2 S_3 
          +a_3^2 S_5\right).
   \end{split}
\end{align}
The unique solution to this equation is
\begin{align}\label{S5}
 S_5\stackrel{a}{=} -\frac{17a_1^2}{3 a_3^2} S_3.
\end{align}
\par

We can use (56') together with \eqref{t2}, \eqref{S4}, \eqref{S1_minus_S2} and
\eqref{S5} to conclude that
\begin{align}
 t_1\stackrel{a}{=}t_2\stackrel{a}{=} 0.
\end{align}
Equation (47') of CE78 tells us that the bounding surface is axisymmetric to the first
PN order if and only if \eqref{S4}, \eqref{S1_minus_S2} and \eqref{S5} hold.
The PN velocity field of CE78 can then be seen to be axisymmetric in the limit we are
discussing, when we additionally require
\begin{align}
 w_2\stackrel{a}{=} -w_1.
\end{align}
\par

Using what has been shown above, the third equation of (58')
in CE78 can be used to find the value of $S_3$ (where the
relationship between the $\alpha^{78}$'s and the $\alpha^{67}$'s can be found
in Appendix~\ref{alpha_relations})
\begin{align}
 0 &= a_1^4Q_{11}^{78} + a_3^4Q_{33}^{78} - a_1^2a_3^2 Q_{31}^{78} \nonumber\\
   &\stackrel{a}{=} a_1^4P_{11}^{78} + a_3^4P_{33}^{78} - a_1^2a_3^2 P_{31}^{78}
        +\frac{130}{9}a_1^2a_3^2A_3S_3\nonumber\\
   &= a_1^4\alpha_{11}^{78} + a_3^4\alpha_{33}^{78} - a_1^2a_3^2 \alpha_{31}^{78} \nonumber\\
    &\quad +\frac{a_1^4Q_1}{4}(q+r_1)+\frac{130}{9}a_1^2a_3^2A_3S_3\nonumber \\
    &\quad +\sum_{i=3}^5 S_i\left(a_1^4u_{11}^{(i)} + a_3^4u_{33}^{(i)} 
          - a_1^2a_3^2 u_{31}^{(i)} \right)\nonumber\\
   \begin{split}
   &= a_1^4\alpha_{11}^{67} + a_3^4\alpha_{33}^{67} - a_1^2a_3^2 \alpha_{31}^{67}\\
    &\quad +\frac{a_1^4Q_1}{2}r_1+\frac{130}{9}a_1^2a_3^2A_3S_3\\
    &\quad +\sum_{i=3}^5 S_i\left(a_1^4u_{11}^{(i)} + a_3^4u_{33}^{(i)} 
          - a_1^2a_3^2 u_{31}^{(i)} \right).
   \end{split}
\end{align}
The solution for $S_3$ is then
\begin{align}
 S_3\stackrel{a}{\approx} -0.01742648312+0.1061462885\,r_1,
\end{align}
the analytic expression of which can be found in Appendix~\eqref{Si_analytic}.

We now turn to the fifth of equations (58') to solve for $S_1$.
The equation is
\begin{align}
 0 &= a_3^4Q_{33}^{78} - a_1^4Q_{11}^{78} + a_3^2 \bar Q_3^{78}- a_1^2 \bar Q_1^{78} \nonumber\\
   &\stackrel{a}{=} a_3^4P_{33}^{78} - a_1^4P_{11}^{78} + a_3^2 P_3^{78}
           - a_1^2P_1^{78}\nonumber\\
     &\quad+2a_1^2a_3^2\left(3S_1+\tfrac{35}{9}S_3\right) \nonumber\\
   \begin{split}
    &= a_3^4\alpha_{33}^{67} - a_1^4\alpha_{11}^{67} + a_3^2 \alpha_3^{67}
         -a_1^2 \alpha_1^{67}-\frac{a_1^4Q_1}{2}r_1\\
     &\quad+2a_1^2a_3^2\left(3S_1+\tfrac{35}{9}S_3\right)- a_1^4 Q_1 w_1\\
     &\quad+ \sum_{i=1}^2 a_1^2 S_i\left(a_3^2 u_3^{(i)} -a_1^2 u_1^{(i)}\right)\\
     &\quad +\sum_{i=3}^5 S_i\left(a_3^4u_{33}^{(i)} - a_1^4u_{11}^{(i)} 
          + a_3^2 u_{3}^{(i)}- a_1^2 u_{1}^{(i)} \right) 
   \end{split}
\end{align}
and the solution is
\begin{align}
 \begin{split}
  S_1&\stackrel{a}{\approx} -(0.2836731908 + 0.7419729757\,r_1\\
     &\quad + 1.121542227\, w_1),
 \end{split}
\end{align}
cf.\ Appendix~\eqref{Si_analytic} for the analytic expression.
The fourth equation is then identically fulfilled. We have obtained a solution
to all the equations at the point $a_2=a_1$ and have two remaining constants,
$w_1$ and $r_1$ (although $q$ and $q_1$ are not determined, they always
appear in the combination $q+q_1$, which is equal to $r_1$, cf.\ \eqref{q_plus_q1}).

\section{The Axisymmetric Limit of a Generalization to Chandrasekhar and Elbert's Paper}
\label{axisymm_limit}

Before discussing the solution obtained above, we consider the solution to the
post-Newtonian equations not at the point $a_2=a_1$, but in the {\sl limit} $a_2\to a_1$.
The equations listed above, are also obtained as limiting relations.
However, in the limit, we also obtain two
new equations, one of which allows us to determine $\lim_{a_2\to a_1} q_1$
and the other, say $\lim_{a_2\to a_1} r_1$.
\par

Equations (24'), (28') and (38') of CE78 provide a system of three linear equations
for the quantities $q_1$, $q_2$ and $q_3$. After solving this linear system, the
limit $a_2\to a_1$ can be taken to give
\begin{align}
 \begin{split}
 q_1&\to -6\sqrt{2B_{11}}\left(4a_1^2B_{111}+\frac{a_1^4}{a_3^2}B_{113} \right)\\
    &= \frac{-\left(2e^2+1\right)^2Q_1^3}{e^4}\\
   &\approx 2.827158725,
 \end{split}
\end{align}
where we have defined the eccentricity
\begin{align}
 e:=\sqrt{1-a_3^2/a_1^2}
\end{align}
and where the explicit expression for $Q_1$ is
\begin{align}
 \lim_{a_2\to a_1}Q_1=\lim_{a_2\to a_1}-\sqrt{\frac{8e^2(1-e^2)}{3+8e^2-8e^4}}
\end{align}
(we remind the reader that $a_3$ depends on $a_1$ and $a_2$ via \eqref{a3_equation}).
\par

The fourth of equations (58') is identically fulfilled for $a_2=a_1$.
Therefore, we introduce
\begin{align}
 \varepsilon:=1-a_2^2/a_1^2
\end{align}
and expand the quantities involved and
solve to first order in $\varepsilon$ to give
\begin{align}\label{r_1}
 \lim_{a_2\to a_1}r_1\approx 0.02880590648-1.75\lim_{a_2\to a_1}w_1.
\end{align}
We provide the analytic expression in Appendix~\ref{Si_analytic}.
\par

Strictly speaking, we have to show that the fourth of equations (58') is
fulfilled to all orders in $\varepsilon$ to be certain that \eqref{r_1}
is continuously connected to the PN Dedekind solutions. We
were able to solve the whole system of equations along the post-Newtonian
Dedekind sequence for arbitrary $w_1$ and $w_2$, meaning that the limit
presented here can be tacked on continuously.

\section{Discussion}\label{discussion}

The axially symmetric post-Newtonian solutions we have generated
depend on two parameters or one if we require that the solution
be continuously connected to the PN Dedekind `ellipsoids'
with the velocity field \eqref{velocity}. The solutions are not
uniformly rotating in general. If we add this constraint, then
requiring that the four-velocity be shear-free tells us that
\begin{align}
 r_1\to 0\quad \text{(shear free)}
\end{align}
must hold.
\par

We now show that with this additional constraint,
the solution is indeed the PN
Maclaurin solution (thereby demonstrating that the shear free condition
is not only necessary, but also sufficient for uniform
rotation in our case). Let us first note that upon taking into
account the results above and in particular $Q_2\to-Q_1$, the
components of the velocity become
\begin{align}
\begin{split}
 v_1 &= \phantom{-}\sqrt{\pi G\rho}\left(Q_1+\frac{\pi G\rho}{c^2}a_1^2 w_1\right)x_2\\
 v_2 &= -\sqrt{\pi G\rho}\left(Q_1+\frac{\pi G\rho}{c^2}a_1^2 w_1\right)x_1\\
 v_3 &= 0.
\end{split}
\end{align}
This is precisely the form of the velocity for the post-Newtonian Maclaurin
spheroids, as can be found in \citet{Chandrasekhar67_C67} (C67a) equation
(3), where $\Omega$ is a constant containing a Newtonian and post-Newtonian
contribution, cf.\ (28) of that paper.
\par

Next we note, that for a given equation of state, an axially symmetric, stationary
and uniformly rotating fluid is described by two parameters.
For our purposes, we can take them to be $a_3/a_1$, which we prescribe
using \eqref{a3_equation},  and the value for $a_1$, which we leave
undetermined.
\par

One has two additional degrees of freedom, which amount to the mapping between a
Newtonian and post-Newtonian solution and is a matter of convention
\citep[cf.][]{Bardeen71}. For
example, one can write the coordinate volume of the star to be
\begin{align}
 V = V_0 + V_1\delta + \ldots,
\end{align}
where $\delta$ is some relativistic parameter, and then {\sl choose}
to have the PN contribution vanish, $V_1=0$. This is the choice that
was made in CE78 and C67b and also in Chandrasekhar's original paper
on the PN Maclaurin spheroids C67a. We have followed this covention
in the current paper, making it easy to compare our results to those
of C67a. The second degree of freedom one has was left unspecified in much
of C67a, though Table~I lists values with the choice
$S_1^\text{M}=S_3^\text{M}=0$.%
\footnote{Where necessary, we distinguish the constants of C67a from
those used here by adding the superscript `M'.}
\par

If we introduce the new coordinate
\begin{align}
 \varpi^2:=x_1^2+x_2^2,
\end{align}
and make use of \eqref{S4}, \eqref{S1_minus_S2} and \eqref{S5},
then the bounding surface (cf.\ (47) in CE78) is given by
\begin{align}
 \begin{split}
  0 &= \frac{\varpi^2}{a_1^2} + \frac{x_3^2}{a_3^2} - 1 - \frac{2\pi G\rho}{c^2}
       \Biggl\{S_1\left(\varpi^2-\frac{2a_1^2 x_3^2}{a_3^2}\right)\\
     &\quad + S_3\biggl[\frac53\left(\varpi^2-\frac{a_1^2 x_3^2}{a_3^2}\right) 
            - \frac43\frac{\varpi^4}{a_1^2} + 4\varpi^2\frac{x_3^2}{a_3^2}\\
     &\quad -\frac{17}{9}\frac{a_1^2 x_3^4}{a_3^4}+\frac53x_1^2\left(\frac{\varpi^2}{a_1^2} + \frac{x_3^2}{a_3^2}
            - 1\right)\biggr]\Biggr\}.
 \end{split}
\end{align}
Using the equation for the surface $\varpi^2/a_1^2=1-x_3^2/a_3^2$, which holds at the Newtonian level
and can thus be inserted into the PN term above, one sees that the term with $x_1^2$ vanishes and
one finds that the equation is identical to (42) of C67a if
\begin{align}
 S_3 &= -\frac{9}{13}S_2^\text{M}+\frac{3a_3^2}{13a_1^2}S_3^\text{M}
            \qquad \text{and}\label{S2_M}\\
 S_1 &= S_1^\text{M} + \frac{16}{13}S_2^\text{M} - \frac{a_3^2}{13a_1^2}S_3^\text{M}\label{S1_M}
\end{align}
hold. As mentioned in that paper, $S_3^\text{M}=0$ may be chosen without loss of
generality%
\footnote{Note that \eqref{S2_M} and \eqref{S1_M} together with \eqref{S1_minus_S2}
are equivalent to the three equations (123) of C67b as can be seen either by
taking $S_3^\text{M}=0$ or identifying $\alpha$ of that equation with 
$S_1^\text{M}+\frac{a_3^2}{3a_1^2}S_3^\text{M}$
and $\beta$ with $S_2^\text{M}-\frac{a_3^2}{3a_1^2}S_3^\text{M}$.}
which then leads to a unique relationship between $S_3$ and
$S_2^\text{M}$, which is shown to be correct in Appendix~\ref{S3_analytic}.
The constant $S_1^\text{M}$ can be chosen arbitrarily just as with $S_1$ (which depends on $w_1$).
\par

If one considers the {\sl limit} $a_2\to a_1$ and simultaneously requires that
the star rotate uniformly, then \eqref{r_1} provides
the unique value for $w_1$,
\begin{align}
 w_1\approx 0.01646051799,
\end{align}
which is equivalent to making a choice for
$S_1^\text{M}$ different from the one made in C67a, but no more
and no less physically meaningful.
\par

The most significant result of the analysis of the axisymmetric limit
is that \eqref{r_1} shows us that the rigidly rotating limit ($r_1=0$) and
the original choice of velocity field in CE78 ($w_1=w_2=0$) are
incompatible. While it is possible with that velocity field to find
the post-Newtonian Maclaurin solution {\sl at} the bifurcation point, this
solution is not continuously connected to any other solution. When
considering the question of the existence or non-existence of non-axially
symmetric but stationary solutions, it seems important to retain the possibility
of studying a neighbourhood of the axially symmetric and uniformly rotating limit,
especially since such solutions are known to exist%
\footnote{As far as we know, there exists no formal proof demonstrating
the existence of such solutions. Steps in that direction were taken
by \cite{Heilig95} and the existence has been demonstrated by many groups
that are able to solve Einstein's equations numerically to extremely high
accuracy, see e.g.\ \cite{AKM03b}}.
This possibility was excluded by the approach taken in CE78.
\par

In a follow-up paper, we intend to tackle the problem with a more
general approach that lends itself better to proceeding to
higher post-Newtonian orders, is not as restrictive in the solutions
it permits and allows one to show that the singularity discussed in
CE78 is an artefact of the specific method chosen and not an
inherent property of the post-Newtonian Dedekind solutions \citep[cf.][]{GP10}.

\acknowledgments

We gratefully acknowledge helpful discussions with M.\ Ansorg, J.\ Bi\v c\'ak, J.\ Friedman and R.\ Meinel.
The first author was financially supported by the grants GAUK 116-10/258025 and GACR 205/09/H033 and the second
by the Deutsche Forschungsgemeinschaft as part of the project ``Gravitational Wave Astronomy'' (SFB/TR7--B1).

\appendix

\section{A Detailed Discussion of Chandrasekhar and Elbert's Work}\label{details}
We mentioned in footnote~\ref{different_solution} that we have been unable to reproduce
the values from Table~1' in CE78 that result from solving (58') nor have we succeeded
in finding the source of the discrepancy. It is important to rule out
an error in our understanding of that paper or an error in our own solutions to the equations
presented there, and we therefore provide a detailed discussion here (in this section
we use the velocity field in that paper, i.e.\ $w_1=w_2=0$).
\par

The calculations we performed were done with the aid of computer algebra. As a
test, we did all the calculations using both Maple and Mathematica.
To be absolutely certain that we solved the equations correctly, we
wrote down the line element and energy-momentum tensor as given in
\citet{Chandrasekhar65}, had Mathematica (TTC package) determine
Einstein's equations to first post-Newtonian order and then verified
that they are indeed fulfilled. When the values from Table~1' of CE78
are inserted, then one finds that the condition that pressure vanish
on the surface is violated at a level three orders of magnitude higher
than with the values from our Table~\ref{constants}. We also verified that the
violation vanishes in our case as more significant figures are added.
\par

The solutions we found for $q_1$, $q_2$ and $q_3$ agree with those given in Table~1'
of CE78. This provides strong evidence suggesting that our numerical evaluation
of $a_3/a_1$ for a given $a_2/a_1$ and of the index symbols is correct. 
Moreover, the dependence of $q$, $r_1$, $r_2$, $t_1$ and $t_2$ on $S_i$ as
given in equations (37') and (53')--(56') can be seen to hold both in Table~1'
and Table~\ref{constants}. This indicates strongly that the typo in equation (53')
of CE78 mentioned in footnote~\ref{footnote_typo} is truly only that and that the quantities in the integrability condition of (11') are treated correctly in both papers, leaving only $\delta U$ and $\Phi$ to be verified. 
\par

The system of linear equations providing the values for $S_i$, i.e.\ (58'),
can of course be written as follows:
\begin{align}\label{matrix_equation}
 \begin{pmatrix}
  M_{11} & \cdots & M_{15} \\
  \vdots & \ddots & \vdots  \\
  M_{51} & \cdots & M_{55}
 \end{pmatrix}
 \begin{pmatrix}
  S_1\\ \vdots\\ S_5
 \end{pmatrix} =
 \begin{pmatrix}
  N_1\\ \vdots\\ N_5
 \end{pmatrix}.
\end{align}
For a given value of $a_2/a_1$, the matrix $(M_{ij})$ depends on the $u$'s from
C67b and  via their $S_i$ dependence, indirectly on
$q$, $r_1$, $r_2$, $t_1$ and $t_2$. The vector $(N_i)$ depends on the $\alpha$'s
and again on the (non-$S_i$ dependent part of) $q$, $r_1$, $r_2$, $t_1$ and $t_2$.
We return to a discussion of this equation after mentioning a few incongruities
in CE78.
\par

In (44') a factor $1/(\pi G \rho)$ is missing in $\delta U$ because the equation
is copied directly from (74) of C67b, whereas the relationship
between $p/\rho$ and $\delta U$ is not the same in (39') of CE78 and (75) of C67b.
This mistake is corrected in (45') and (46') however. In (39') there is also
a factor $1/(\pi G \rho)^2$ missing in the term $2\Phi+2v^2U
+\frac12\left(\frac{p}{\rho}\right)^2$ as can be seen by checking dimensions%
\footnote{We advise the reader that, as mentioned after
(14) in \cite{CE74}, the units in which $Q_1$ and $Q_2$ are measured change as of this
point by a factor $\sqrt{\pi G\rho}$.}
and comparing to (11) in \cite{CE74}. Finally, we note that \eqref{matrix_equation} from above
only ensures that the pressure is constant on the surface of the PN-ellipsoid
as discussed in C67b, cf.\ (75) in {\it loc.\ cit.},
but not that it vanishes. The constant that would have to be determined to
ensure vanishing pressure was not written in (39') or (40')%
\footnote{The constant contained in $\delta U$ is completely determined
by (44') and is thus not available as a variable to ensure that the
pressure vanish on the surface.}
and the constant that is a part of $\delta U$ in (44') was dropped when
proceeding to (45'). Since the determination of this constant plays no
role in the paper however, we need not discuss it further and have not
done so in our own paper.
\par

We find that the determinant of $(M_{ij})$ vanishes at $a_2/a_1=0.33700003168\ldots$
just as in CE78, where it is given to four significant figures. This provides evidence
suggesting that the matrices agree (and thus the $\delta U$) and that the vectors $(N_i)$ disagree.
If we multiply $\delta U$ by a factor $\pi$, as suggested in the last paragraph, then the determinant becomes zero for $a_2/a_1=0.30874\ldots$. Nonetheless, we tested that neither an arbitrary factor
in front of this term, nor one in front of the term
$2\Phi+2v^2U+\frac12\left(\frac{p}{\rho}\right)^2$ can explain the results in CE78.
\par

A natural explanation for a disagreement between the vectors $(N_i)$ in our
case and in CE78 would be that one of the $\alpha$'s contains a mistake. We
checked to see that an arbitrary change in a single $\alpha$ cannot account
for the differences in the results however. Since an explicit expression for
these $\alpha$'s is not provided in CE78, we cannot test directly to see whether or not
each agrees. However, in the implicit expressions from (39') and (40'), only the
contributions from $2\Phi+2v^2 U+\frac12\left(\frac{p}{\rho}\right)^2$ are not
written out. These can easily be compared to those written out explicitly
for the $\alpha$'s of C67b, where the appropriate modifications for the
different Newtonian velocity have to be taken into account, and show perfect
agreement with our expressions. In particular, the relationship to the $\alpha$'s of C67b for $a_2=a_1$,
which is discussed in Appendix~\ref{alpha_relations} provides additional evidence for the correctness of our expressions.
We also generated the $\alpha$'s with computer algebra by typing out
the expressions for (11'), solving the integrability condition and integrating it
and showed that these agree with the expressions provided below.%
\footnote{For the terms in (11'), we checked our expressions by ensuring that
$\nabla^2 U=-4\pi G\rho$, (8') and the Newtonian equations hold. Furthermore,
we tested the $u$'s by first ensuring that the moments $\mathfrak{D}_i$,
$\mathfrak{D}_{ijk}$ fulfil the appropriate Poisson equation and that the
$\delta U^{(i)}$ of (69) from C67b agree with (70) and (71) from the same
paper.}
\begin{align}
 \begin{split}
 \alpha_1^{78} &= -\frac{a_3^4}{a_1^2}A_3^2 -4a_2^2B_{12}(A_1+A_2)-2IQ_1Q_2
  -(2I+3a_3^2A_3)A_1\\
  &\phantom{=} + (a_1^2A_{11}-\tfrac12B_{11})(2a_1^2Q_2^2-2a_1^2A_1-3a_3^2A_3)
    - \tfrac12 B_{12}(2a_2^2Q_1^2-2a_2^2A_2-3a_3^2 A_3)+\tfrac52a_3^2A_3B_{13}
 \end{split}\\
 \begin{split}
 \alpha_2^{78} &= -\frac{a_3^4}{a_2^2}A_3^2 -4a_1^2B_{12}(A_1+A_2)-2IQ_1Q_2
  -(2I+3a_3^2A_3)A_2\\
  &\phantom{=} + (a_2^2 A_{22}-\tfrac12 B_{22})(2a_2^2 Q_1^2-2a_2^2 A_2-3a_3^2A_3)
    - \tfrac12 B_{12}(2a_1^2Q_2^2-2a_1^2A_1-3a_3^2 A_3)+\tfrac52a_3^2A_3B_{23}
 \end{split}\\
 \begin{split}
 \alpha_3^{78} &= -a_3^2 A_3^2  -(2I+3a_3^2A_3)A_3
   -\tfrac12 B_{23}(2a_2^2 Q_1^2-2a_2^2 A_2-3a_3^2A_3)\\
  &\phantom{=}  - \tfrac12 B_{13}(2a_1^2Q_2^2-2a_1^2A_1-3a_3^2 A_3)+\tfrac52a_3^2A_3(B_{33}-2a_3^2A_{33})
 \end{split}\\
 \begin{split}
 \alpha_{12}^{78} &= \frac{a_3^4}{a_1^2a_2^2}A_3^2 -2Q_1^2\left(A_1+\frac{a_2^4}{a_1^4}A_2\right)
   + (2a_1^2Q_2^2-2a_1^2A_1-3a_3^2 A_3)(-a_1^2A_{112}+\tfrac12B_{112})\\
  &\phantom{=} + (2a_2^2 Q_1^2-2a_2^2 A_2-3a_3^2A_3)(-a_2^2A_{122}+\tfrac12B_{122})
   - \tfrac52a_3^2A_3B_{123} + 2Q_1Q_2\left(1-\frac{a_2^2}{a_1^2}\right)A_2\\
  &\phantom{=} - \tfrac12Q_1^3Q_2
   - 2a_2^2Q_1Q_2(3A_{22}+A_{12}) + 4Q_1^2\left(A_1-\frac{a_2^2}{a_1^2}A_2\right)-Q_1(q_1+\tfrac12q_2)
 \end{split}\\
 \begin{split}
 \alpha_{23}^{78} &= \frac{a_3^2}{a_2^2}A_3^2 -2Q_1^2A_3 + 2Q_1^2\left(1-\frac{a_2^2}{a_1^2}\right)A_3
   - 2a_1^2Q_1Q_2(A_{13}+A_{23}) + (2a_1^2Q_2^2-2a_1^2A_1-3a_3^2 A_3)(\tfrac12B_{123})\\
  &\phantom{=} + (2a_2^2 Q_1^2-2a_2^2 A_2-3a_3^2A_3)(-a_2^2A_{223}+\tfrac12B_{223})
   - \tfrac52a_3^2A_3(-2a_3^2A_{233}+B_{233}) 
 \end{split}\\
 \begin{split}
 \alpha_{31}^{78} &= \frac{a_3^2}{a_1^2}A_3^2 -2Q_2^2A_3 + 2Q_1Q_2\left(1-\frac{a_2^2}{a_1^2}\right)A_3
   - 2a_2^2Q_1Q_2(A_{13}+A_{23}) - \tfrac52a_3^2A_3(-2a_3^2A_{133}+B_{133})\\
  &\phantom{=} + (2a_1^2Q_2^2-2a_1^2A_1-3a_3^2 A_3)(-a_1^2A_{113}+\tfrac12B_{113})
               + (2a_2^2 Q_1^2-2a_2^2 A_2-3a_3^2A_3)(\tfrac12B_{123}) 
 \end{split}\\
 \begin{split}
 \alpha_{11}^{78} &= \tfrac12\frac{a_3^4}{a_1^4} A_3^2 -2Q_2^2 A_1
   + (2a_1^2Q_2^2-2a_1^2A_1-3a_3^2 A_3)(-a_1^2A_{111}+\tfrac14B_{111})\\
  &\phantom{=} + (2a_2^2 Q_1^2-2a_2^2 A_2-3a_3^2A_3)(\tfrac14B_{112})
   - \tfrac54a_3^2A_3B_{113} + Q_1Q_2\left(1-\frac{a_2^2}{a_1^2}\right)A_1\\
  &\phantom{=} - \tfrac14Q_1Q_2^3
   - a_2^2Q_1Q_2(3A_{11}+A_{12}) -\tfrac14Q_2q_1
 \end{split}\\
 \begin{split}
 \alpha_{22}^{78} &= \tfrac12\frac{a_3^4}{a_2^4} A_3^2 -2Q_1^2 A_2
   + (2a_1^2Q_2^2-2a_1^2A_1-3a_3^2 A_3)(\tfrac14B_{122})\\
  &\phantom{=} + (2a_2^2 Q_1^2-2a_2^2 A_2-3a_3^2A_3)(-a_2^2A_{222}+\tfrac14B_{222})
   - \tfrac54a_3^2A_3B_{223} + Q_1^2\left(1-\frac{a_2^2}{a_1^2}\right)A_2\\
  &\phantom{=} - \tfrac14Q_1^3Q_2
   - Q_1Q_2a_1^2(3A_{22}+A_{12}) -\tfrac14Q_1q_2
 \end{split}\\
 \begin{split}
 \alpha_{33}^{78} &= \tfrac12 A_3^2
   + \tfrac14(2a_1^2Q_2^2-2a_1^2A_1-3a_3^2 A_3)B_{133}
   + \tfrac14(2a_2^2 Q_1^2-2a_2^2 A_2-3a_3^2A_3)B_{233}\\
  &\phantom{=} - \tfrac52a_3^2A_3(-2a_3^2A_{333}+\tfrac12B_{333})
 \end{split}
\end{align}
\par

Let us summarize the arguments from above. We have checked all the equations in 
Part~I of CE78 and find the analytic expressions to be free of error, except for
the few minor points mentioned above. We have good reason to believe that both
in that paper and here, Einstein's PN-equations are solved correctly including
the PN-Bianchi identity. We obtain different numerical values for $S_i$ which
we suspect is related to a problem with the numerical evaluation of the $\alpha$'s
in CE78, though we cannot be certain that our matrices $(M_{ij})$ agree simply
because their determinants vanish at the same point. The various tests of our
$\alpha$'s and the fact that we find the
post-Newtonian Maclaurin spheroids in the axisymmetric case convince us that
our values are correct.

\begin{deluxetable}{cccccccccccccc}
\tabletypesize{\footnotesize}
\rotate
\tablecaption{The numerical values we find for the quantities listed in Table~1 of \citet{CE78_CE78}.\label{constants}}
\tablewidth{0pt}
\tablehead{
\colhead{$a_2/a_1$} & \colhead{$q_1$} & \colhead{$q_2$} & \colhead{$q_3$} & \colhead{$S_1$} & \colhead{$S_2$} & \colhead{$S_3$} & \colhead{$S_4$} & \colhead{$S_5$} & \colhead{$q$} & \colhead{$r_1$} & \colhead{$r_2$} & \colhead{$t_1$} & \colhead{$t_2$}
}
\startdata
1.00 & 2.8272 & -2.8272 & 0.0000 & -0.3050 & -0.3290 & -0.0144 & 0.0574 & 0.2398 & -2.7984 & 0.0288 & -0.0288 & 0.0000 & 0.0000\\
0.99 & 2.8173 & -2.8370 & 0.0211 & -0.2944 & -0.3323 & -0.0132 & 0.0542 & 0.2481 & -2.7984 & 0.0196 & -0.0378 & -0.0454 & -0.0445\\
0.98 & 2.8073 & -2.8470 & 0.0424 & -0.2838 & -0.3355 & -0.0120 & 0.0508 & 0.2565 & -2.7984 & 0.0101 & -0.0466 & -0.0923 & -0.0887\\
0.97 & 2.7972 & -2.8570 & 0.0639 & -0.2733 & -0.3387 & -0.0107 & 0.0474 & 0.2648 & -2.7984 & 0.0002 & -0.0553 & -0.1407 & -0.1324\\
0.96 & 2.7869 & -2.8671 & 0.0857 & -0.2628 & -0.3416 & -0.0093 & 0.0438 & 0.2732 & -2.7984 & -0.0101 & -0.0638 & -0.1906 & -0.1757\\
0.95 & 2.7766 & -2.8774 & 0.1077 & -0.2524 & -0.3445 & -0.0078 & 0.0401 & 0.2815 & -2.7985 & -0.0207 & -0.0722 & -0.2422 & -0.2186\\
0.90 & 2.7231 & -2.9297 & 0.2213 & -0.2010 & -0.3573 & 0.0010 & 0.0200 & 0.3223 & -2.7989 & -0.0804 & -0.1121 & -0.5287 & -0.4282\\
0.85 & 2.6665 & -2.9843 & 0.3414 & -0.1509 & -0.3675 & 0.0121 & -0.0030 & 0.3610 & -2.7999 & -0.1528 & -0.1486 & -0.8741 & -0.6316\\
0.80 & 2.6067 & -3.0412 & 0.4691 & -0.1019 & -0.3754 & 0.0258 & -0.0288 & 0.3949 & -2.8020 & -0.2413 & -0.1818 & -1.2996 & -0.8317\\
0.75 & 2.5439 & -3.1006 & 0.6052 & -0.0538 & -0.3813 & 0.0426 & -0.0572 & 0.4198 & -2.8064 & -0.3506 & -0.2120 & -1.8363 & -1.0329\\
0.70 & 2.4781 & -3.1627 & 0.7511 & -0.0064 & -0.3856 & 0.0636 & -0.0871 & 0.4280 & -2.8152 & -0.4872 & -0.2398 & -2.5324 & -1.2409\\
0.66 & 2.4236 & -3.2146 & 0.8758 & 0.0313 & -0.3882 & 0.0845 & -0.1106 & 0.4131 & -2.8280 & -0.6222 & -0.2610 & -3.2547 & -1.4177\\
0.65 & 2.4098 & -3.2279 & 0.9082 & 0.0408 & -0.3888 & 0.0905 & -0.1161 & 0.4049 & -2.8324 & -0.6603 & -0.2662 & -3.4656 & -1.4642\\
0.64 & 2.3958 & -3.2414 & 0.9412 & 0.0503 & -0.3895 & 0.0968 & -0.1215 & 0.3944 & -2.8374 & -0.7005 & -0.2715 & -3.6910 & -1.5118\\
0.63 & 2.3818 & -3.2550 & 0.9747 & 0.0598 & -0.3901 & 0.1037 & -0.1265 & 0.3814 & -2.8432 & -0.7427 & -0.2768 & -3.9323 & -1.5607\\
0.62 & 2.3677 & -3.2687 & 1.0087 & 0.0693 & -0.3907 & 0.1109 & -0.1311 & 0.3653 & -2.8497 & -0.7873 & -0.2821 & -4.1913 & -1.6111\\
0.61 & 2.3536 & -3.2826 & 1.0434 & 0.0789 & -0.3914 & 0.1187 & -0.1352 & 0.3459 & -2.8573 & -0.8343 & -0.2876 & -4.4697 & -1.6632\\
0.60 & 2.3394 & -3.2967 & 1.0786 & 0.0886 & -0.3921 & 0.1271 & -0.1388 & 0.3225 & -2.8659 & -0.8840 & -0.2931 & -4.7697 & -1.7171\\
0.59 & 2.3252 & -3.3109 & 1.1145 & 0.0983 & -0.3929 & 0.1362 & -0.1416 & 0.2947 & -2.8758 & -0.9367 & -0.2989 & -5.0939 & -1.7732\\
0.58 & 2.3109 & -3.3254 & 1.1510 & 0.1082 & -0.3938 & 0.1460 & -0.1436 & 0.2616 & -2.8872 & -0.9924 & -0.3048 & -5.4449 & -1.8317\\
0.57 & 2.2967 & -3.3400 & 1.1881 & 0.1182 & -0.3948 & 0.1567 & -0.1444 & 0.2224 & -2.9004 & -1.0516 & -0.3110 & -5.8262 & -1.8929\\
0.56 & 2.2824 & -3.3548 & 1.2260 & 0.1284 & -0.3960 & 0.1685 & -0.1439 & 0.1763 & -2.9155 & -1.1146 & -0.3175 & -6.2416 & -1.9574\\
0.55 & 2.2681 & -3.3699 & 1.2646 & 0.1387 & -0.3973 & 0.1814 & -0.1418 & 0.1219 & -2.9330 & -1.1818 & -0.3244 & -6.6957 & -2.0254\\
0.50 & 2.1975 & -3.4487 & 1.4690 & 0.1956 & -0.4087 & 0.2720 & -0.0912 & -0.3361 & -3.0751 & -1.5988 & -0.3691 & -9.7775 & -2.4444\\
0.45 & 2.1301 & -3.5353 & 1.6959 & 0.2722 & -0.4377 & 0.4537 & 0.1171 & -1.4497 & -3.4139 & -2.2489 & -0.4521 & -15.4156 & -3.1217\\
0.40 & 2.0694 & -3.6327 & 1.9502 & 0.4253 & -0.5330 & 0.9625 & 0.9405 & -4.8952 & -4.4794 & -3.5956 & -0.6789 & -29.1295 & -4.6607\\
0.35 & 2.0208 & -3.7456 & 2.2386 & 1.6155 & -1.4902 & 5.7347 & 10.0089 & -37.9506 & -15.1294 & -13.4666 & -2.7757 & -137.5070 & -16.8446\\
0.34 & 2.0132 & -3.7707 & 2.3012 & 6.5850 & -5.5917 & 26.0552 & 49.3174 & -178.2659 & -60.8513 & -54.2058 & -11.6728 & -582.0542 & -67.2855\\
0.33 & 2.0065 & -3.7968 & 2.3655 & -2.6453 & 2.0383 & -11.7342 & -23.8765 & 82.4680 & 24.2357 & 21.4267 & 4.8654 & 241.2314 & 26.2701\\
0.32 & 2.0007 & -3.8240 & 2.4317 & -1.0183 & 0.6975 & -5.0896 & -11.0416 & 36.4961 & 9.3003 & 8.0964 & 1.9536 & 94.8022 & 9.7077\\
0.30 & 1.9925 & -3.8819 & 2.5703 & -0.4078 & 0.1986 & -2.6145 & -6.2981 & 19.0909 & 3.7752 & 3.1511 & 0.8615 & 37.3393 & 3.3605\\
0.28 & 1.9892 & -3.9456 & 2.7179 & -0.2340 & 0.0578 & -1.9180 & -4.9749 & 13.7746 & 2.2579 & 1.8969 & 0.5449 & 17.7005 & 1.3877\\
0.25 & 1.9960 & -4.0540 & 2.9584 & -0.1417 & -0.0205 & -1.5433 & -4.1994 & 9.8393 & 1.4994 & 1.8182 & 0.3549 & 0.1194 & 0.0075\\
\enddata
\end{deluxetable}

\subsection{The Solution at the Bifurcation Point}\label{alpha_relations}
At the point $a_2=a_1$, i.e.\ at the bifurcation point along the Maclaurin sequence,
the following relations can be used to simplify the expressions for the $\alpha$'s,
where $\Omega$ refers to the angular velocity of the uniformly rotating Newtonian
solution and has the same meaning as in C67b:
\begin{align}
 a_3^2 A_3=a_1^4 A_{11}=a_1^2(A_1-B_{11})=I-2a_1^2A_1=a_1^2\left(A_1-\tfrac12 Q_1^2\right), \qquad \Omega^2=2B_{11}=Q_1^2.
\end{align}
Note that at this point, the $\alpha$'s of C67b and C67a agree and we find
\begin{align}
 \begin{split}
 \alpha_1^{78} &= -\frac{a_3^4}{a_1^2}A_3^2 -8a_1^2 B_{11}A_1 + 2IQ_1^2 -(2I+3a_3^2A_3)A_1
   + (a_1^2A_{11}- B_{11})(2a_1^2Q_1^2-2a_1^2A_1-3a_3^2A_3)
    +\tfrac52a_3^2A_3B_{13}\\
  &= -15a_1^2A_1^2-\tfrac{19}{4}a_1^2 Q_1^4 +14a_1^2A_1Q_1^2 + \tfrac52a_3^2A_3B_{13}\\
  &= \alpha_1^{67} = \alpha_2^{78} = \alpha_2^{67}
 \end{split}\\
 \begin{split}
 \alpha_3^{78} &=  -a_3^2 A_3^2  -(2I+3a_3^2A_3)A_3
   - B_{13}(2a_1^2Q_1^2-2a_1^2A_1-3a_3^2A_3) +\tfrac52a_3^2A_3(B_{33}-2a_3^2A_{33})\\
  &= a_1^2(2Q_1^2-10A_1)A_3 + a_1^4Q_1^2A_{13}\\
  &= \alpha_3^{67}
 \end{split}\\
 \begin{split}
 \alpha_{12}^{78} &= \frac{a_3^4}{a_1^4} A_3^2 -4Q_1^2A_1
   + a_1^2(\tfrac72Q_1^2-5A_1)(-2a_1^2A_{111}+B_{111})- \tfrac52a_3^2A_3B_{113} + \tfrac12Q_1^4
   + 8a_1^2Q_1^2A_{11} -\tfrac12 q_1 Q_1\\
     &= \alpha_{12}^{67}-\tfrac12 q_1 Q_1
 \end{split}\\
 \begin{split}
 \alpha_{23}^{78} &= \frac{a_3^2}{a_1^2}A_3^2 -2Q_1^2A_3 + 4a_1^2Q_1^2A_{13} 
  + a_1^2(\tfrac72Q_1^2-5A_1)(B_{113}-a_1^2A_{113})- \tfrac52a_3^2A_3(-2a_3^2A_{133}+B_{133})\\
  &=\alpha_{23}^{67}=\alpha_{31}^{78}=\alpha_{31}^{67}
 \end{split}\\
 \alpha_{11}^{78} &= \alpha_{22}^{78} =\tfrac12\alpha_{12}^{78}+\tfrac12Q_1q_1=\alpha_{11}^{67}+\tfrac14Q_1q_1\\
 \begin{split}
 \alpha_{33}^{78} &= \tfrac12A_3^2 
  + \tfrac12 a_1^2(\tfrac72Q_1^2-5A_1)B_{133}- \tfrac52a_3^2A_3(-2a_3^2A_{333}+\tfrac12B_{333})\\
  &=\alpha_{33}^{67}
 \end{split}
\end{align}

\section{Explicit expressions for $S_1$, $S_3$ and $r_1$}\label{Si_analytic}

At the point $a_2= a_1$, the $u$'s from C67b and C67a are
related by
\begin{align}
 u_{ij}^{(2)\text{M}} = \left.-\frac{9}{13}\left(u_{ij}^{(3)}-4u_{ij}^{(4)}
  -\frac{17a_1^2}{3a_3^2}u_{ij}^{(5)}\right)\right|_{a_2=a_1},
\end{align}
which follows from (119) of C67b.
\par

Using these relations, those between the $\alpha$'s and \eqref{q_minus_r1}, \eqref{S4}
and \eqref{S5}, one finds that the third of equations (58) of CE78 becomes
\begin{align}
 0 &= a_1^4Q_{11}^{78} - a_1^2a_3^2Q_{13}^{78} + a_3^4Q_{33}^{78}\nonumber\\
   &\stackrel{a}{=} a_1^4\alpha_{11}^{67}-a_1^2a_3^2\alpha_{13}^{67}+a_3^4\alpha_{33}^{67}
      + \frac{a_1^4Q_1}{2}r_1 +\frac{130}{9}a_1^2 a_3^2A_3 S_3 
      -\frac{13}{9}\left(a_1^4u_{11}^{(2)\text{M}}-
                  a_1^2a_3^2u_{13}^{(2)\text{M}}+a_3^4u_{33}^{(2)\text{M}}\right)S_3
\end{align}
We thus have the solution
\begin{align}\label{S3_analytic}
 S_3 = \frac{9}{13}
  \frac{a_1^4\alpha_{11}^{67}-a_1^2a_3^2\alpha_{13}^{67}+a_3^4\alpha_{33}^{67}
           +a_1^4Q_1r_1/2}
  {a_1^4u_{11}^{(2)\text{M}}
    -a_1^2a_3^2u_{13}^{(2)\text{M}}+a_3^4u_{33}^{(2)\text{M}}-10a_1^2a_3^2A_3},
\end{align}
which agrees with (99) of C67a if we take \eqref{S2_M} of this paper into account.
In order to provide concise explicit formul\ae, we again make use of the eccentricity
\begin{align*}
 e=\sqrt{1-a_3^2/a_1^2},
\end{align*}
the quantity
\begin{align}
 C:=104e^6-444e^4+630e^2-245
\end{align}
and recall that  $Q_1$ is
\begin{align*}
 Q_1\stackrel{a}{=} -\sqrt{\frac{8e^2(1-e^2)}{3+8e^2-8e^4}}.
\end{align*}
We now provide explicit expressions for $S_1$, $S_3$ and $r_1$. Note that
the expressions for $S_1$ and $S_3$ can be obtained either as limiting values
or by placing oneself directly on the point $a_2=a_1$. On the other hand, $r_1$
can only be obtained by a limiting process. The formul\ae\ read
\begin{align}
 \begin{split}
 S_1&\stackrel{a}{=}
      \frac{e}{2e^2-1}\biggl[\frac{-1}{26eC}(2864e^8-10128e^6+14712e^4-8120e^2+1365)Q_1^2\\
     &\quad      +\frac{e}{3Q_1}w_1 + \frac{4e}{39CQ_1}(224e^6-840e^4+1170e^2-455)r_1\biggr],
 \end{split}\\
 S_3&\stackrel{a}{=}
    \frac{36e^4}{65C}\left[\frac{(272e^4-244e^2+35)Q_1^2}{8e^2}-\frac{3e^2}{Q_1}r_1\right],\\
  r_1&\to \frac{-Q_1^3}{8e^2(2e^2+1)}(24e^4-12e^2-1)-\frac74w_1.
\end{align}
In deriving these expressions, we have made use of the identities
\begin{align}
 \left. a_3^2\left(4A_{11}-\frac{2}{a_1^2}\right)-4a_1^2 A_{11}+3A_1\right|_{a_2=a_1}=0,\\
 \left. 3A_1^2-3A_1-4a_1^2 A_1A_{11}+5a_1^2 A_{11}-2a_1^4 A_{11}^2\right|_{a_2=a_1}=0.
\end{align}

\bibliographystyle{apj}
\bibliography{Reflink}

\end{document}